\documentclass{PoS}
\usepackage{epsfig}
\PoS{PoS(LAT2005)168}

\title{Series representation; Pade' approximants and critical behavior in QCD at nonzero T and mu }

\ShortTitle{Series representation for QCD at nonzero T and mu }

\author{\speaker{Maria Paola Lombardo} \\  
       Laboratori di Frascati dell' 
       Istituto Nazionale di Fisica Nucleare, I-00044 Frascati(RM), Italy \\
        E-mail: \email{lombardo@lnf.infn.it}}

\abstract{We discuss the  analytic continuation 
beyond $\mu/T \simeq 1$   in QCD at nonzero T and $\mu$
by use of the Pade' approximants. The slope of the critical line 
obtained in this way increases at 
large $\mu$  with respect to the second order Taylor result.In the
hot phase Pade' and Taylor approximants coincide, suggesting a very
large, and possibly infinite, radius of convergence of the Taylor series
in this thermodynamic region. }

\FullConference{XXIIIrd International Symposium on Lattice Field Theory\\
25-30 July 2005\\
Trinity College, Dublin, Ireland}

\begin{document}
\section{Introduction}

Results obtained at imaginary $\mu_B$ can be analytically continued to
real $\mu_B$\cite{OP,Lombardo:1999cz,deForcrand:2002ci,D'Elia:2002gd}. 
In principle,  rigorous arguments guarantee that the 
analytic continuation 
of a function can be  done within the entire analytic domain.
In practice, the exact analytic form is not known, and a systematic procedure
relying on the Taylor expansion is only valid within the circle of
convergence of the series itself.
In this talk we discuss how to implement
the analytic continuation of the critical line and
of thermodynamics observables beyond the circle of convergence
of the Taylor series in a controlled way.

Let us remind ourselves that an analytic function is locally
      representable as  a Taylor series. The convergence  disks can be
      chosen is such a way that they overlap two by two, and cover the
      analytic   domain.  Thus,   one  way   to  build   the  analytic
      continuation   is  by  connecting   all  of   these  convergence
      disks. The arcs of the  convergence circles which are within the
      region where f is analytic have a pure geometric meaning, and by
      no means are an obstacle to the analytic continuation.
      Assume now that the  circle of convergence about $z$  = (0,0) has
      radius  unit, i.e.  is  tangent  to the  lines  which limit  the
      analytic domain; take now a $z$ value, say $z_1 = (0,a), 1/2 < a < 1  $
      inside the convergence  disk as  the origin  of a  new series
      expansion, which is explicitly defined by the rearrangement 
                  $(z- z_0)^n = (z - z_1 + z_1 - z_0)^n$ 
      As the  radius of  convergence of the  new series will  be again
      one, this procedure will extend  the domain of definition of our
      original  function (the  two series  define restrictions  of the
      same function to the intersection between the two disks), and by
      'sliding'  the convergence disk  we can  cover all  the analytic
      strip.

We have sketched above the standard theoretical argument
to demonstrate the feasibility of analytic continuation beyond the
radius of convergence, and we will show that 
the Pade' series is one practical way to accomplish it.

\section{Naive, but maybe useful examples}

Let us consider a function $f(z)$ of a complex variable $z$ which is analytic
within a strip limited by $z= \pm i$, and let us assume that the 
radius of convergence of the Taylor expansion about $z=0$ is 
given by the distance
from the nearest singularity (it could be larger, see discussions below).
This is the kind of pattern expected for temperatures above 
the end point of the chiral transition line, the nearest line of 
singularities being associated to the Roberge--Weiss transition.
A possible function realizing this pattern is
$f1(z) = \frac {1}{z^2 + 1}$. For real $z = 1$ the Taylor approximants 
the partial sums $S_n$ are  $S_n = 1$ for $n$ even, and $S_n = 2$
for $n$ odd, demonstrating the lack of convergence of the series. 
On the other hand, obviously, the Pade' approximants
$P[N,M]$ not only converge, but become exact as soon as $M > 1$
on the entire real axes. Similar exercises can be repeated for less trivial 
example, for instance the function 
$f2(z) = \frac {e^{-z}}{z^2 + 1}$
is well represented  by the Pade' approximants P[N,N] for N > 2 
well beyond the convergence disk.

\section{ The Critical Line beyond $\mu/T \simeq 1$}

We can measure the critical line for imaginary $\mu$ ,
and our goal is to analytically continue the results on the right hand
side. The radius of convergence of the Taylor representation of
the critical line might well be limited by the Roberge Weiss singularities.
However, as explained before, the Pade' approximation is not.

In Fig. 1 we present the Pade' analysis of data for
four \cite {D'Elia:2002gd} and two flavor \cite{deForcrand:2002ci}.
\begin{figure}[t]
{\epsfig{file=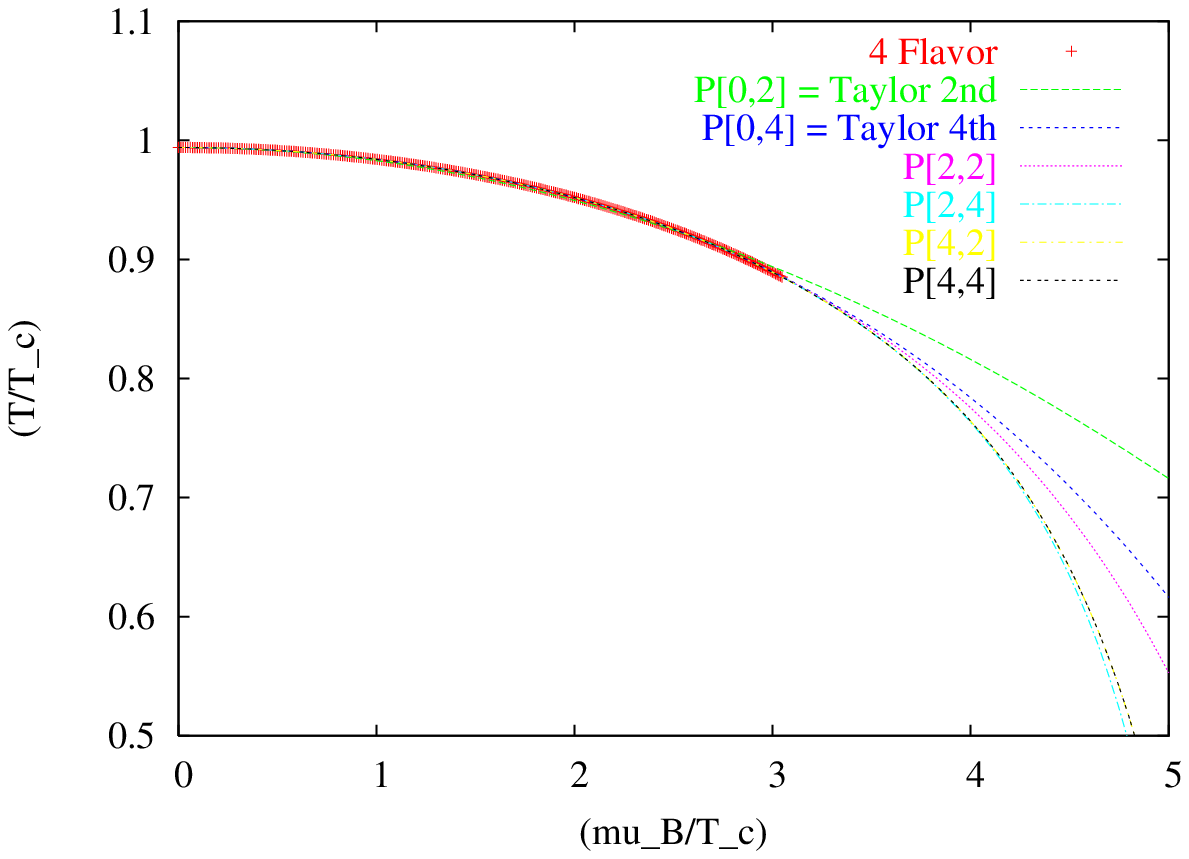, width = 11 truecm}} \\ 
{\epsfig{file=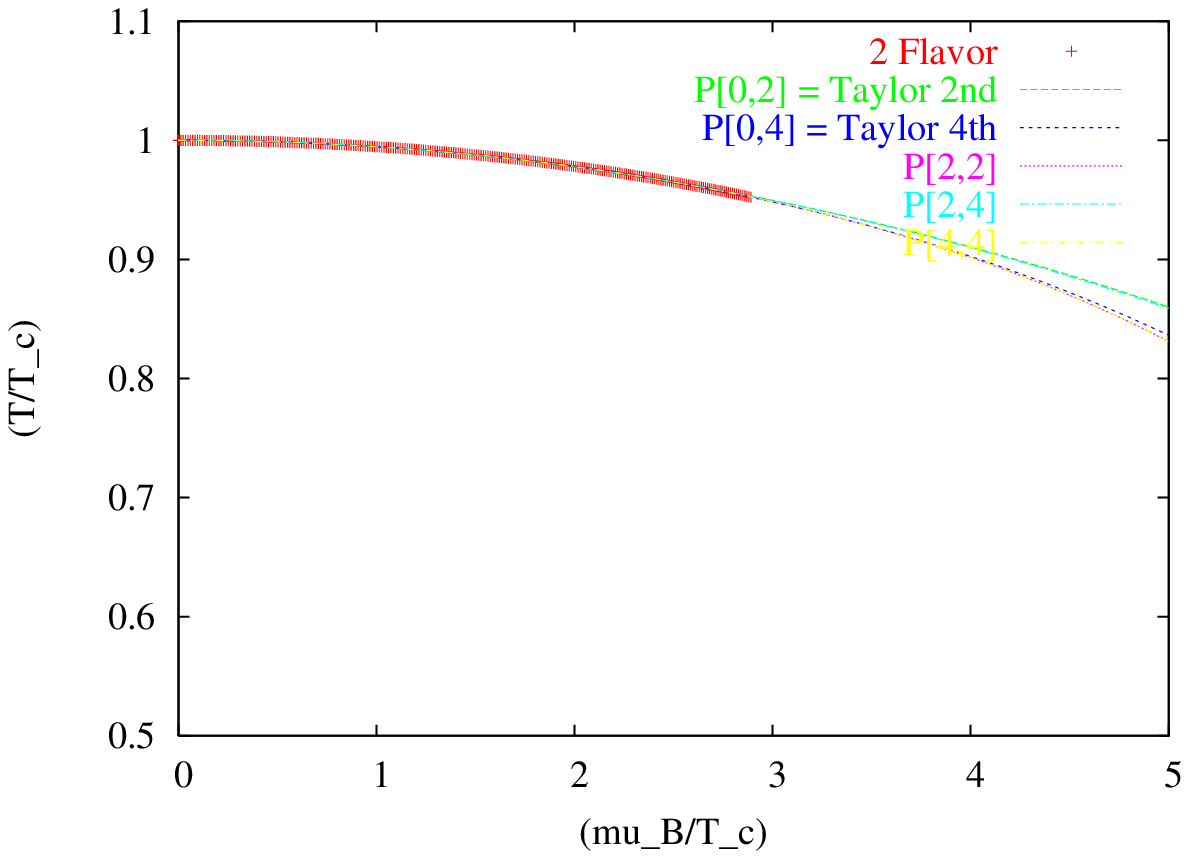, width = 11 truecm}}
\caption{Pade' approximants for the critical line of four 
(top) and two flavor QCD. The red, bold line extends up to the radius
of convergence of the Taylor series $\mu/T \simeq 1$, 
and has been fitted to the Pade' approximants ,  which  converge well  
beyond it.}
\end{figure}
Results seem stable beyond $\mu_B = 500 MeV (\mu_B/T \simeq 1)$, 
with the Pade' analysis  in good agreement with Taylor expansion
for smaller $\mu$ values. At larger $\mu$ the Taylor expansion 
seems less stable, while the Pade' still converges, giving a slope 
of the critical line larger than the naive continuation of the
second order Taylor approximations. The same behavior is suggested by
recent results within the canonical approach\cite {kra} 
and the DOS method \cite{chr}.

We underscore that the possibility of analytically continue the
results beyond the radius of convergence of the Taylor series
by no means imply that one can blindly extrapolate a lower order
approximation! Even when it is \underline{possible}
to achieve convergence -- via Pade' approximants, or within the
convergence radius of the Taylor series --  one has always to cross 
check different orders of approximation to make sure that convergence has
indeed been achieved. For instance, it would then be interesting to repeat 
the comparisons between the second order  results 
shown in \cite{Azcoiti:2005tv} by extending the Taylor series to
fourth order and/or by use of Pade' approximants.
\begin{figure}[t]
{\epsfig {file=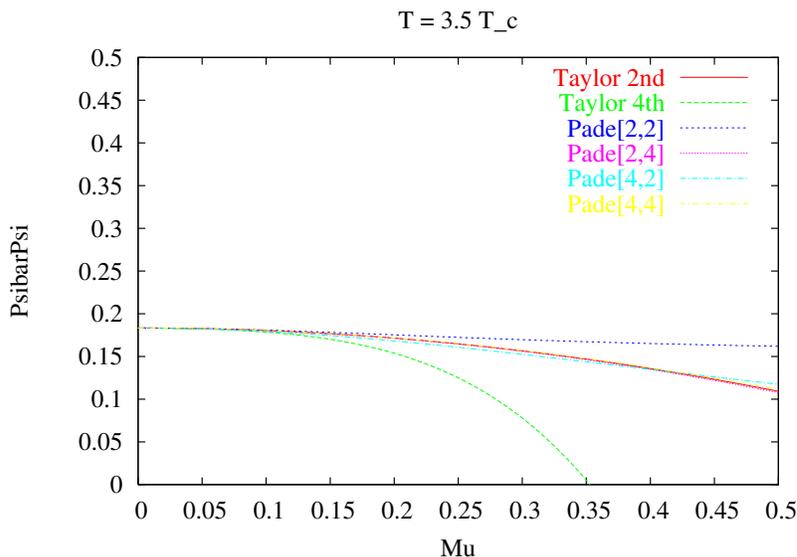, width = 11 truecm}}
\caption{Analytic continuation via Pade' and Taylor approximants for 
the chiral condensate in the hot phase for $T = 3.5 T_c$: 
note that both series seem
to converge to the same result, suggesting a very large radius of convergence
of the Taylor series in this thermodynamic region.}
\end{figure}
\begin{figure}
{\epsfig{file=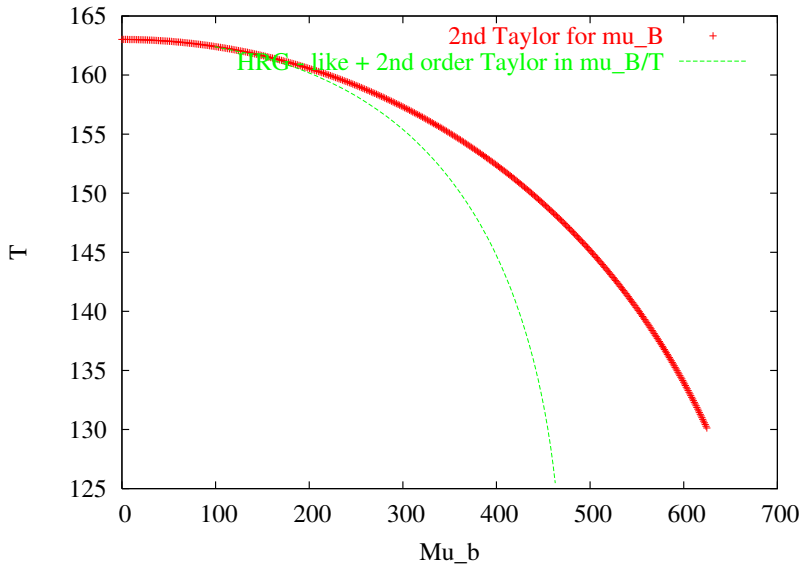, width = 11 truecm}}
\caption{The critical line of the four flavor model computed
by use of the simplified parametrization of the HRG model discussed
in the text.}
\end{figure}
\section{Interlude -- Radius of convergence and critical behavior}

I mentioned en passant that the radius of convergence of a Taylor
expansion about the origin might well be larger than the distance 
of the origin itself from the
nearest singularity. While complex analysis textbooks offer a full discussion
of this point, I would like to single out here three cases which might be
encountered in usual critical behavior:
$f1(z) = A1(z) (1 - z/z_c)^{-\lambda} ;
f2(z) =  A2(z) \theta(z -z_c) ;
f3(z)  =  A3(z) \theta(z -z_c)(1 - z/z^*)^{-\lambda}$ , where $An(z)$
is an analytic function.

Case 1 corresponds to an usual critical behavior (second order or larger).
Case 2 represents a strong first order phase transition. Case 3 is
intermediate between the two, a weak first order transition at $z_c$,
and a spinodal point at $z^*$. Correspondingly, we have  
different radius of convergence of the  Taylor series: 
$r_1 = |z_c| , r_2 = \infty, r_3 = z^*$ : in conclusion
the radius of convergence of the Taylor expansion
for the critical line and thermodynamics observables 
 might be  infinite as well a finite, depending on the nature of the 
Roberge Weiss transition. Conversely, if the nature of the phase transition 
is known, one can infer from it the radius of convergence of the Taylor
series, as done by Gavai and Gupta \cite{Gavai:2004sd}, which in turn
locates the critical point.

\section{ Beyond $\mu/T \simeq 1$  in the Hot Phase}

The Pade' approximants to the results for the chiral
condensate in the hot phase are shown 
in Fig. 2, where I used four flavor data \cite{D'Elia:2004at}.

Again we see that the  Pade' analysis 
seems capable to produce stable results. We should also note that
the Taylor expansion seems stable as well, which might indicate a large 
(infinite?) radius of convergence in this range of temperature.
Indeed, as noted in \cite{D'Elia:2004at} the radius of convergence 
should tend to infinite in the infinite
temperature limit, and indeed it has been estimated to be large 
by the Bielefeld--Swansea collaboration\cite{Allton:2005gk}.
A detailed investigation at imaginary $\mu$ of the region closer to 
the critical temperature is in progress \cite{qcdatwork}. 

\section { The critical line from the Hadron Gas}

An alternative way to analytically continue the
results relies on phenomenological modeling.
The   Hadron Resonance Gas model
might provide a description 
of QCD thermodynamics in the confined, hadronic phase  of QCD
\cite{Karsch:2003zq,Allton:2005gk,D'Elia:2004at},
and can be used to determine the critical line as well.

The critical temperature as a function of
$\mu_B$ is determined by lines of constant energy
density: $\epsilon \simeq 0.5 - 1.0 $~GeV/fm$^3$\cite {Toublan:2004ks}
A continuation of the critical line using the HRG ansatz
plus a fixed energy (or any other quantity determined at $\mu = 0$ )
criterion suggests the implicit form for the critical line
$
T = f(T) cosh(\mu_B/T)
$ with $\lim_{\mu_B/T \to 0}  f(T) cosh (\mu_B/T) = 1-  k \mu^2$.
We have naively approximated $f(T) =  1-  { k} \mu^2$, and used
the resulting form to fit the data in the $\mu/B < 1$ range.
Accrdoing to the above discussion, this again can be continued
beyond this limit, and   also in this case we get a critical line whose
slope increases with increasing $\mu$ (Fig. 3).

\section{ Summary}

For $T > T_c$ Pade' approximants are a viable tool to analytically continue
thermodynamic results obtained at imaginary chemical potential to
the entire analytic domain, beyond the radius of convergence
of the Taylor expansion.

The stability  of the Taylor partial sums for $T \simeq 3.5 T_c$ 
hints at an infinite radius of convergence of the Taylor series
at high temperature, according with a naive expectations of a  
strong first order  Roberge Weiss  transition at high T .

For $T < Tc$  
the  analytic continuation is best performed via a Fourier series
which converge (one term alone) to the Hadron Resonance Gas model.
In turn, the HRG parametrisation might be used to constrain the
analytic form of the critical line.

Pade' analysis,  and an HRG parametrisation of
the critical line, are viable tools to continue the critical
line beyond $\mu/T \simeq  1$.  The results suggest that the 
curvature of the critical line increases at lower temperature.

\section*{Acknowledgments}
This research was supported in part by the National Science Foundation 
under Grant No. PHY99-07949, and I wish to thank the participants in
the KITP program 'Modern Challenges for Lattice QCD' for helpful 
conversations.


\begin{thebibliography}{99}
\bibitem{OP} O. Philipsen, this volume.
\bibitem{Lombardo:1999cz}
  M.~P.~Lombardo,
  Nucl.\ Phys.\ Proc.\ Suppl.\  {\bf 83} (2000) 375.
\bibitem{deForcrand:2002ci}
  P.~de Forcrand and O.~Philipsen,
  Nucl.\ Phys.\ B {\bf 642} (2002) 290.
\bibitem{D'Elia:2002gd}
  M.~D'Elia and M.~P.~Lombardo,
  Phys.\ Rev.\ D {\bf 67} (2003) 014505
\bibitem{Azcoiti:2005tv}
V.~Azcoiti, G.~Di Carlo, A.~Galante and V.~Laliena,
  Nucl.\ Phys.\ B {\bf 723} (2005) 77; V. Laliena, this volume.
\bibitem{Gavai:2004sd}
  R.~V.~Gavai and S.~Gupta,
  Phys.\ Rev.\ D {\bf 71} (2005) 114014; S. Gupta, this volume.
\bibitem{D'Elia:2004at}
  M.~D'Elia and M.~P.~Lombardo,
  Phys.\ Rev.\ D {\bf 70} (2004) 074509
\bibitem{Allton:2005gk}
  C.~R.~Allton {\it et al.},
  Phys.\ Rev.\ D {\bf 71} (2005) 054508 and references therein.

\bibitem{qcdatwork} M. D'Elia, F. Di Renzo, M.P. Lombardo, 
arXiv:hep-lat/0511029, and work in progress.
\bibitem{kra} Ph. de Forcrand and S. Kratochvila, this volume
\bibitem{chr} C. Schmidt, Z. Fodor, S. Katz, this volume

\bibitem{Karsch:2003zq}
  F.~Karsch, K.~Redlich and A.~Tawfik,
  Phys.\ Lett.\ B {\bf 571} (2003) 67

\bibitem{Toublan:2004ks}
  D.~Toublan and J.~B.~Kogut,
  Phys.\ Lett.\ B {\bf 605} (2005) 129.

\end{thebibliography}
\end{document}